\documentclass[american,aps,pra,reprint,superscriptaddress]{revtex4-1}

\usepackage[T1]{fontenc}
\usepackage[sc]{mathpazo}
\usepackage{amsmath}
\usepackage{amssymb}
\usepackage{enumerate}
\usepackage{amsthm}
\usepackage{color}

\newtheorem{lem}{Lemma}

\usepackage{amsfonts,mathrsfs}
\usepackage[style]{fncychap}
\usepackage{graphicx} 
\usepackage{geometry} 
\usepackage{eepic}
\usepackage{ifthen}
\newboolean{ElectronicVersion}
\setboolean{ElectronicVersion}{true} 

\usepackage[T1]{fontenc}
\usepackage[sc]{mathpazo}
\usepackage{mathdots}
\usepackage{amsmath}
\usepackage{amssymb}
\usepackage{enumerate}
\usepackage{amsthm}
\usepackage{amsfonts,mathrsfs}
\usepackage[style]{fncychap}
\usepackage{graphicx} 
\usepackage{geometry} 
\usepackage{eepic}
\usepackage{ifthen}
\usepackage{cases}

\usepackage{amsfonts,mathrsfs}
\usepackage[style]{fncychap}
\usepackage{graphicx} 
\usepackage{geometry} 
\usepackage{eepic}
\usepackage{ifthen}
\usepackage{cases}

\usepackage[unicode=true,pdfusetitle, bookmarks=true,bookmarksnumbered=false,bookmarksopen=false, breaklinks=false,pdfborder={0 0 0},backref=false,colorlinks=false] {hyperref}
\hypersetup{ colorlinks,linkcolor=myurlcolor,citecolor=myurlcolor,urlcolor=myurlcolor}
\definecolor{myurlcolor}{rgb}{0,0,0.7}

\def\be{\begin{equation}}
\def\ee{\end{equation}}
\def\bea{\begin{eqnarray*}}
\def\eea{\end{eqnarray*}}

\def\I{\mathbb{1}}

\newenvironment{mylist}[1]{\begin{list}{}{
    \setlength{\leftmargin}{#1}
    \setlength{\rightmargin}{0mm}
    \setlength{\labelsep}{2mm}
    \setlength{\labelwidth}{8mm}
    \setlength{\itemsep}{0mm}}}
    {\end{list}}

\newcommand{\bra}[1]{\langle#1|}

\newcommand{\ket}[1]{|#1\rangle}

\newcommand{\braket}[1]{\langle#1\rangle}

\def\H{\textsf{H}}

\def\X{\textsf{X}}

\newtheorem{thrm}{Theorem}

\theoremstyle{definition}

\newtheorem{remark}{Remark}

\numberwithin{equation}{section}
\def\be{\begin{equation}}
\def\ee{\end{equation}}
\def\bea{\begin{equation*}}
\def\eea{\end{equation*}}
\def\bna{\begin{eqnarray*}}
\def\ena{\end{eqnarray*}}
\def\bn{\begin{eqnarray}}
\def\en{\end{eqnarray}}
\def\bpm{\begin{pmatrix}}
\def\epm{\end{pmatrix}}

\newcounter{questionnumber}

\usepackage{chngcntr}
\counterwithout{equation}{section}

\begin{document}

\title{Embedding, simulation and consistency of $\cal PT$ -symmetric quantum Theory}

 \author{Minyi Huang}
 \email{11335001@zju.edu.cn}
 \affiliation{School of Mathematical Sciences, Zhejiang University, Hangzhou 310027, PR~China}

 \author{Asutosh Kumar}
 \email{Corresponding author: asutoshk.phys@gmail.com}
 \affiliation{P.G. Department of Physics, Gaya College, Magadh University, Rampur, Gaya 823001, India}

 \author{Junde Wu}
 \email{Corresponding author: wjd@zju.edu.cn}
\affiliation{School of Mathematical Sciences, Zhejiang University, Hangzhou 310027, PR~China}

\begin{abstract}
A physical requirement on the Hamiltonian operator in quantum mechanics is that it must generate real energy spectrum and unitary time evolution. While the Hamiltonians are Dirac Hermitian in conventional quantum mechanics, they observe $\cal PT$-symmetry in $\cal PT$-symmetric quantum theory.
The embedding property was first studied by G\"unther and Samsonov [\href{https://journals.aps.org/prl/abstract/10.1103/PhysRevLett.101.230404}{Phys. Rev. Lett. {\bf 101}, 230404 (2008)}] to visualise the evolution of unbroken $\cal PT$-symmetric Hamiltonians on $\mathbb C^2$ by Hermitian Hamiltonians on $\mathbb C^4$.
This paper investigates the properties of $\cal PT$-symmetric quantum systems including the embedding property. We provide a full characterization of the embedding property in the general case and show that only unbroken $\cal PT$-symmetric quantum systems admit this property in a finite dimensional space. Furthermore, utilizing this property, we establish a physically realizable simulation process of the unbroken $\cal PT$-symmetric Hamiltonians. An observation that the unbroken $\cal PT$-symmetric quantum systems can be viewed as open systems in the conventional quantum mechanics accounts for the consistency of $\cal PT$-symmetric quantum theory.
\end{abstract}

\maketitle

\section{Introduction}
Quantum mechanics is the most successful theory of nature at the microscopic scale so far. In
this theory, a fundamental physical requirement on the Hamiltonian operator (of a closed quantum-mechanical system) is that it must generate real energy spectrum and unitary time evolution. In conventional quantum mechanics, the Hamiltonian operator is considered Dirac Hermitian to ensure that the energy spectrum is real and that the time evolution is unitary (that is, probability conserving). To this end, it is interesting to ask what if one considers a complex non-Hermitian Hamiltonian? It was believed that such a Hamiltonian could not describe a valid quantum-mechanical theory because the non-Hermiticity of the Hamiltonian would result in nonunitary time evolution \cite{nonunitary1, nonunitary2, nonunitary3, nonunitary4}. However, note that the requirement of Hermiticity is an axiom. It might be possible to replace the condition of Hermiticity by some more general property without losing
any of the essential physical features of quantum mechanics. Such a possibility was investigated by Bender and Boettcher \cite{bender1998real} in 1998. In fact, they found numerically that the eigenvalues of the complex non-Hermitian Hamiltonians of the form, ${\cal H} = \hat{p}^2 - (i\hat{x})^N$, with $N$ being a continuous parameter are entirely
real and positive for all $N \geq 2$, while for $N<2$ the spectrum is partly real and partly complex. These Hamiltonians are
$\cal PT$-symmetric, i.e., they observe space-time reflection symmetry. In $\cal PT$-symmetric quantum theory, Hamiltonians are $\cal PT$-symmetric. It should be noted that the condition of Hermiticity is not wrong but also not necessary. $\cal PT$-symmetric quantum theory is merely a complex generalization of conventional quantum theory.
Theories defined by non-Hermitian $\cal PT$-symmetric Hamiltonians exhibit strange and unexpected properties
both at the classical and at the quantum level \cite{bender1998real}.
Moreover, Mostafazadeh has studied the general pseudo-Hermitian theory \cite{mostafazadeh2010pseudo}.
$\cal PT$-symmetric theory has been successfully applied to quantum optics and quantum
statistical mechanics, etc. \cite{ruter2010observation,PhysRevA.82.013629,PhysRevA.87.051601,chang2014parity, deffner2015jarzynski}. Very recently,  $\cal PT$-symmetry has been revisited to study the Riemann hypothesis \cite{bender17}.

However, $\cal PT$-symmetry theory has offered certain hitches now and then.
Some proposed applications of $\cal PT$-symmetry theory in the quantum information problems such as the quantum brachistochrone problem \cite{bender2007faster, PhysRevA.78.042115}, quantum state discrimination \cite{bender2013pt}, and increasing entanglement by local $\cal PT$-symmetric operations \cite{PhysRevA.90.054301} have been argued to be somewhat controversial.
In fact, it was believed that a probabilistic simulation of $\cal PT$-symmetric Hamiltonians is needed for experimental realization of such applications. We address this problem in this paper.
In such a scenario, the embedding property is useful. This property was first studied by G\"unther and Samsonov \cite{gunther2008naimark} to visualise the evolution of unbroken $\cal PT$-symmetric Hamiltonians on $\mathbb C^2$ by Hermitian Hamiltonians on $\mathbb C^4$.
%
It was also noted that the $\cal PT$-symmetric theory might violate the no-signaling principle \cite{lee2014local}. Thus, it arose the consistency problem with the conventional quantum mechanics.
To deal with this issue, Croke and Brody proposed several schemes in which the consistency of $\cal PT$-symmetry theory was usually argued from the viewpoint of closed systems \cite{croke2015pt,brody2016consistency}. A recent experimental investigation \cite{tang2016experimental} of the $\cal PT$-symmetric theory has confirmed a formal (although not essential!) violation of the no-signaling condition. Such a result can be analysed only in the framework of open systems.
We will see that the embedding property takes into account this point.

In this paper, we investigate the properties of $\cal PT$-symmetric quantum systems including the embedding property.
Firstly, we give a full characterization of the embedding property. Based on this characterization, a simulation paradigm of unbroken $\cal PT$-symmetric Hamiltonians is presented. This provides a general method to realize a $\cal PT$-symmetric quantum system and helps quantitatively analyse the effect of $\cal PT$-symmetric Hamiltonians. Our simulation paradigm addresses the consistency issue from the viewpoint of open systems.
The paper is organised as follows. In section II, we briefly review $\cal PT$-symmetry, and discuss the condition for unbroken $\cal PT$-symmetric operator. In section III, we provide a full characterization of the embedding property. We present a physically realizable simulation paradigm of the unbroken $\cal PT$-symmetric Hamiltonians in section IV. In section V, the simulation paradigm is called in to address the consistency problem. Finally, the conclusion is presented in section VI.

\section{Preliminaries}
Throughout this paper, we consider only finite dimensional complex Hilbert space $\mathbb{C}^n$ in which we denote the elements by column vectors, the set of $n\times n$ complex matrices by $M_n(\mathbb C)$, the identity matrix of order $n$ by $I_n$ or simply by $I$, and the complex conjugation of $z\in\mathbb{C}$ and $A\in M_n(\mathbb C)$ by $\overline{z}$ and $\overline{A}$ respectively. $A^\dag$ is the transpose of $\overline{A}$.
An operator $\cal A$ on $\mathbb{C}^n$ is said to be linear if ${\cal A}(c_1x_{1}+c_2x_{2})=c_1{\cal A}(x_{1})+c_2{\cal A}(x_2)$ and anti-linear if ${\cal A}(c_1x_{1}+c_2x_{2})=\overline{c_1}{\cal A}(x_{1})+\overline{c_2}{\cal A}(x_2)$ for $c_i\in\mathbb{C}$, $x_i\in \mathbb{C}^n$, $i=1, 2$.
Let $\{e_i\}_{i=1}^n$ be a basis, $\cal P$ be a linear operator and $\cal T$ be an anti-linear operator on
$\mathbb{C}^n$. Then
${\cal P}(e_j)=\sum_kP_{kj}e_k ~~ \text{and} ~~ {\cal T}(e_j)=\sum_kT_{kj}e_k$,
where $P=(P_{kj})$ and $T=(T_{kj})$ are said to be the representation matrices of $\cal
P$ and $\cal T$ respectively.
If $\{P_i\}_{i=1, 2}$ and $\{T_i\}_{i=1,2}$ respectively are the matrices of linear operators $\{{\cal P}_i\}_{i=1, 2}$ and anti-linear operarors $\{{\cal T}_i\}_{i=1, 2}$ then the representation matrices of
${\cal P}_1{\cal P}_2$, ${\cal T}_1{\cal T}_2$, ${\cal T}_1{\cal
P}_2$ and ${\cal P}_1{\cal T}_2$ are $P_1P_2$, $T_1\overline{T_2}$,
$T_1\overline{P_2}$ and $P_1T_2$ respectively \cite{uhlmann16}.

\subsection{Parity ($\cal P$), Time reversal ($\cal T$) and $\cal PT$-symmetric Hamiltonian ($\cal H$)}
A linear operator $\cal P$ is said to be a parity operator if ${\cal P}^2=\cal I$. And, an anti-linear operator $\cal T$ is said to be a time reversal operator if ${\cal T}^2={\cal I}$. (See also Ref. \cite{note0}.)
%
In physics, there is a natural physical requirement that ${\cal P}$ should commute with ${\cal T}$. This, together with anti-linearity of ${\cal PT}$, promotes ${\cal PT}$ as a time reversal operator.
%
A linear operator $\cal H$ is $\cal PT$-symmetric (that is, $\cal H$ observes space-time reflection symmetry) if $[{\cal H}, {\cal PT}]=0$. Let $P$, $T$ and $H$ respectively be the matrices of ${\cal P}$, ${\cal T}$ and $\cal H$. Then the above conditions reduce to $P^2=I=T\overline{T}$, $PT=T\overline{P}$ and $HPT=PT\overline{H}$.
Note that, since ${\cal PT}$ is an anti-linear operator, a $\cal PT$-symmetric Hamiltonian $\cal H$ is said to be ``unbroken'' if all of the eigenfunctions of $\cal H$ are simultaneously eigenfunctions of $\cal PT$. 
Alternately, for a given $\cal PT$, a linear operator $\cal H$ on $\mathbb{C}^n$ is said to be ``unbroken'' $\cal PT$-symmetric if (i) $\cal H$ is $\cal PT$-symmetric, and (ii) there exist eigenvectors $\{\psi_i\}$ of $\cal H$ such that ${\cal PT}\psi_i=\psi_i$ and $\{\psi_i\}$ spans the whole space $\mathbb{C}^n$. Otherwise, it is said to be ``broken''.
The above condition (ii) can be rephrased in matrix language: (ii') there exists an invertible matrix $\Psi$ such that $\Psi^{-1}H\Psi$ is diagonal and $PT\overline{\Psi}=\Psi$.
Note that if ${\cal H}\psi_i=\lambda_i\psi_i$ and ${\cal PT}\psi_i=\psi_i$, then condition (i) ensures that $\lambda_i$'s are real. Thus, unbroken $\cal PT$-symmetric operators have only real eigenvalues
\cite{bender2007making,mostafazadeh2010pseudo,deng2012general,mannheim2013pt}.

Below we recollect some lemmas 
which are useful for our study.
\begin{lem}\cite{horn2012matrix}
\label{lem1}
If $\cal T$ is a time reversal operator and $T$ is its repersentation matrix then there exists an invertible matrix
$\Psi_1$ such that $T=\Psi_1\overline{\Psi_1}^{-1}$.
\end{lem}

\begin{lem}\cite{horn2012matrix}
\label{lem2}
A matrix $A$ is similar to a real matrix
if and only if there exists an invertible Hermitian matrix $Q$ such that $QAQ^{-1}=A^\dag$.
\end{lem}

\begin{lem}\cite{horn2012matrix}
\label{lem3}
Each real matrix is similar to
\begin{equation}
J=
\begin{pmatrix}
\begin{smallmatrix}
J_{n_1}(\lambda_{n_1},\overline{\lambda}_{n_1})&&&&&\\
&\ddots&&&\\
&&J_{n_p}(\lambda_{n_p},\overline{\lambda}_{n_p})&&&&\\
&&&J_{n_q}(\lambda_{n_q})&&\\
&&&&\ddots&\\
&&&&&J_{n_r}(\lambda_{n_r})
\end{smallmatrix}
\end{pmatrix},
\label{cano2}
\end{equation}
where $J_{n_k}(\lambda_{n_k},\overline{\lambda}_{n_k})=
J_{n_k}(\lambda_{n_k})\oplus J_{n_k}(\overline{\lambda_{n_k}})=
\begin{pmatrix}\begin{smallmatrix}J_{n_k}(\lambda_{n_k})&0\\0&J_{n_k}(\overline{\lambda_{n_k}})\end{smallmatrix}\end{pmatrix}$,
$J_{n_j}(\lambda_{n_j})$ is the Jordan block, and $\lambda_{n_q},
\cdots, \lambda_{n_r}$ are real numbers.
\end{lem}

By lemma \ref{lem1}, since ${\cal PT}$ is a time reversal operator, there exists an
invertible matrix $\Psi_1$ such that $PT=\Psi_1\overline{\Psi_1}^{-1}$. Hence, for a $\cal PT$-symmetric operator $\cal H$, we have
\begin{equation}
\Psi_1^{-1}H\Psi_1=\overline{\Psi_1^{-1}H\Psi_1}.
\end{equation}
%
That is, for any $\cal PT$-symmetric $\cal H$, its
representation matrix $H$ is similar to a real matrix. Thus, Lemma
\ref{lem3} gives the canonical form of a $\cal PT$-symmetric
operator $\cal H$.

The following theorem establishes an important relation between $PT$ and $H$.
\begin{thrm}\label{thmhpt}
Let $\cal H$ be a $\cal PT$-symmetric operator and $H$ be its
representation matrix. Then there exists an invertible matrix $\Psi$
such that
\begin{equation*}
\Psi^{-1}H\Psi=J ~~\text{in Lemma \ref{lem3}},
\end{equation*}
and
\[\Psi^{-1} PT\overline{\Psi}=K,
\] where
\begin{equation*}
K=\begin{pmatrix}
\begin{smallmatrix}
\bpm\begin{smallmatrix}
0&I\\I&0\end{smallmatrix}\epm&&&&&\\
&\ddots&&&&\\
&&\bpm\begin{smallmatrix}
0&I\\I&0\end{smallmatrix}\epm&&&\\
&&&I&&\\
&&&&\ddots&\\
&&&&&I
\end{smallmatrix}
\end{pmatrix}.
\label{cano3}
\end{equation*}
\end{thrm}

\begin{proof}
(i) If $H$ is similar to $J_k(\lambda_k)\oplus
J_k(\overline{\lambda_k})$ then there exists a matrix
$\Phi_1=(\psi_1,\cdots,\psi_k,\psi_{k+1},\cdots,\psi_{2k})$ with $\psi_i$'s being column vectors such
that $H\Phi_1=\Phi_1\bpm\begin{smallmatrix} J_k(\lambda_k)&0\\0&
J_k(\overline{\lambda_k})\end{smallmatrix}\epm$. Since
$PT\overline{H}=HPT$ and $H\psi_{i+1}=\lambda\psi_{i+1}+\psi_i$ ($i=1,\cdots, k-1$), one can verify that
$HPT\overline{\psi_{i+1}}=\overline{\lambda}
PT\overline{\psi_{i+1}}+PT\overline{\psi_i}$.
Let $\Psi=(\psi_1,\cdots,\psi_k,PT\overline{\psi_1},\cdots,PT\overline{\psi_k})$. Then we have
$H\Psi=\Psi \bpm \begin{smallmatrix}J_k(\lambda_k)&0\\0& J_k(\overline{\lambda_k})\end{smallmatrix}\epm$ and
$PT\overline{\Psi}=\Psi\bpm\begin{smallmatrix} 0 & I\\I & 0\end{smallmatrix}\epm$.
(ii) If $H$ is similar to $J_k(\lambda_k)$ then, by Lemma \ref{lem1}, $PT=\Phi_2\overline{\Phi_2}^{-1}$ and
$H'=\Phi_2^{-1}H\Phi_2$ is a real matrix. Note that $H$ is similar to $J_k(\lambda)$, so is $H'$. Since $H'$ is real, there exists a real invertible matrix $\Omega_2$ such that $H'\Omega_2=\Omega_2J_k(\lambda_k)$.
It then follows that
\begin{eqnarray*}
&&H\Phi_2\Omega_2=\Phi_2\Omega_2J_k(\lambda),\\
&&PT\overline{\Phi_2\Omega_2}=\Phi_2\Omega_2.
\end{eqnarray*}
The last identity holds because $\overline{\Omega_2}=\Omega_2$. Taking $\Psi=\Phi_2\Omega_2$, we have $\Psi^{-1}H\Psi=J_k(\lambda_k)$ and $\Psi^{-1}PT\overline{\Psi}=I$.

For a general $H$, there exists a $\Psi$ such that $\Psi^{-1}H\Psi$ is $J$ in Lemma \ref{lem3} and $K=\Psi^{-1}PT\overline{\Psi}$ is the direct sum of the two cases above. This completes the proof.
\end{proof}

Theorem \ref{thmhpt} can be equivalently stated in the operator language as follows.

(i) When $H$ is similar to $J_k(\lambda_k)\oplus J_k(\overline{\lambda_k})$, there exist basis vectors
$\{\psi_1,\cdots,\psi_k,{\cal PT}\psi_1,\cdots,{\cal PT}\psi_k\}$ such that
\begin{eqnarray*}
&&({\cal H}-\lambda{\cal I})^{k}\psi_k=0, \\
&&{\cal H}\psi_{i+1}=\lambda\psi_{i+1}+\psi_i,\\
&&{\cal HPT}\psi_{i+1}=\overline{\lambda}{\cal PT}\psi_{i+1}+{\cal PT}\psi_i,
\end{eqnarray*}
where $i=1,\cdots, k-1$.

(ii) When $H$ is similar to $J_k(\lambda)$, there exist basis vectors
$\{\psi_1,\cdots,\psi_k\}$ such that
\begin{eqnarray*}
&&({\cal H}-\lambda{\cal I})^{k}\psi_k=0,\\
&&{\cal H}\psi_{i+1}=\lambda\psi_{i+1}+\psi_i,\\
&&{\cal PT}\psi_k=\psi_k,
\end{eqnarray*}
where $i=1,\cdots, k-1$.

\begin{remark}\label{rk1} Note that Theorem \ref{thmhpt} also implies that any matrix similar to $J$ in Lemma \ref{lem3} corresponds to some $\cal PT$-symmetric operator. To see this, consider a linear operator $\cal H$, with its representation matrix $H$, and a matrix $\Psi$ such that $\Psi^{-1}H\Psi=J$ in Lemma \ref{lem3}. Let $PT=\Psi K\overline{\Psi^{-1}}$, where $K$ is the matrix in Theorem \ref{thmhpt}. Direct calculations show that $PT\overline{PT}=I$ and $PT\overline{H}=HPT$. That is, $\cal H$ is $\cal PT$-symmetric, with respect to $PT=\Psi K\overline{\Psi^{-1}}$. Following the preceding discussions, we learn that a linear operator $\cal H$ is $\cal PT$-symmetric if and only if its representation matrix is similar to the canonical form $J$ in Lemma \ref{lem3}. It should also be noted that since $\Psi$ is not unique, the constructions of $\cal P$ and $\cal T$ are not unique. 
\end{remark}

As a further application of Remark \ref{rk1}, we show that any linear operator $\cal H$ with representation matrix $H$ similar to some real diagonal matrix can be viewed as unbroken $\cal PT$-symmetric. To see this, suppose $\Psi^{-1}H\Psi=J$, where $J$ is now a real diagonal matrix, a special case of Eq. (\ref{cano2}) in which all $J_{n_i}(\lambda_{n_i},\overline{\lambda}_{n_i})$ vanish and all $J_{n_i}(\lambda_{n_i})$ are of order one. Now Theorem \ref{thmhpt} suggests that $K=I$. Hence, we take $PT=\Psi\overline{\Psi^{-1}}$ as was done in Remark \ref{rk1}. One can directly verify that $\cal H$ is unbroken $\cal PT$-symmetric. Thus, combining with the definition of unbroken $\cal PT$-symmetric operators, we have: a linear operator $\cal H$ is unbroken $\cal PT$-symmetric if and only if its representation matrix $H$ is similar to a real diagonal matrix ($\cal H$ is diagonalisable and has only real eigenvalues).



\subsection{Metric operator $\eta$ of $\cal H$}
In quantum mechanics, the time evolution of a quantum system for a given Hamiltonian $\cal H$
must be unitary. Let $\phi_1$ and $\phi_2$ be two quantum states. By Schr\"{o}dinger equation, the final states are given by $e^{-it\cal H}\phi_1$ and $e^{-it\cal H }\phi_2$. If $\cal H$ is
non-Hermitian then $\braket{e^{-it\cal H}\phi_1,e^{-it\cal
H}\phi_2}=\braket{\phi_1,e^{it{\cal H}^{\dagger}}e^{-it\cal
H}\phi_2}\neq\braket{\phi_1,\phi_2}$, which implies that the
evolution is not unitary.
As a remedy to this non-unitary evolution for a non-Hermitian Hamiltonian, one may introduce a
Hermitian operator $\eta$ on $\mathbb{C}^n$ and define an
$\eta$-inner product as
$\braket{\phi_1,\phi_2}_{\eta}=\braket{\phi_1,\eta\phi_2}$. Then
$\cal H$ will give a unitary evolution with respect to the
$\eta$-inner product if and only if ${\cal H}^\dag\eta=\eta \cal H$.
By Lemmas \ref{lem1} and \ref{lem2}, such an $\eta$ always exists.
The operator $\eta$ is said to be a metric operator of $\cal H$
\cite{mostafazadeh2010pseudo,deng2012general,mannheim2013pt,bagarello2016non,horn2012matrix}.
In the rest of this article, we will use $\eta$ for both the metric operator (of $\cal H$) and its matrix representation.
The following theorem associates $\eta$ with $\cal PT$-symmetric operator $\cal H$.

\begin{thrm}\cite{gohberg1983matrices2}
\label{thmheta}
$\eta$ is a metric operator of a given $\cal PT$-symmetric operator $\cal H$
(that is, ${\cal H}^\dag \eta=\eta {\cal H}$), if and only if there exist a matrix
$\Xi$ such that
\begin{equation*}
\Xi^{-1}H\X=J ~~\text{in Lemma \ref{lem3}},
\end{equation*}
and
\begin{equation}
\begin{smallmatrix}
\Xi^\dag\eta\Xi=S=
\begin{pmatrix}
\begin{smallmatrix}
 S_{2n_1}&&&&&\\
&\ddots&&&\\
&& S_{2n_p}&&&&\\
&&&\epsilon_{n_q} S_{n_q}&&\\
&&&&\ddots&\\
&&&&&\epsilon_{n_r} S_{n_r}
\end{smallmatrix}
\end{pmatrix},
\label{cano3}
\end{smallmatrix}
\end{equation}
where $n_i$'s are the orders of the Jordan blocks $J_{n_i}(\lambda_{n_i},\overline{\lambda}_{n_i})$, $S_{k}=\begin{pmatrix}\begin{smallmatrix}&&1\\&\iddots&\\1&&\end{smallmatrix}\end{pmatrix}_{k\times k}$ and $\epsilon_i=\pm 1$ are uniquely determined by $\eta$.
\end{thrm}

One can also write Theorem \ref{thmheta} in the language of operators as follows.

(i) When $H$ is similar to $J_k(\lambda_k)\oplus J_k(\overline{\lambda_k})$, $\braket{\xi_i,\eta\xi_{j}}=\delta_{i,2k-j+1}$,
where $\xi_{k+i}={\cal PT}\xi_i$ ($i=1,\cdots,k$).
(ii) When $H$ is similar to $J_k(\lambda_k)$, $\braket{\xi_i,\eta\xi_j}=\pm\delta_{i,k-j+1}$.

As a special case of Theorem \ref{thmheta}, if $H$ is unbroken (that is, $J$ is real and diagonal) then
$\braket{\xi_i,\eta\xi_j}=\epsilon_i\delta_{ij}$. Moreover, if $\epsilon_i=1$ for all $i$ then the $\eta$-inner product is positive-definite.
On the other hand, the positivity of $\eta$ and $S$ in Eq. (\ref{cano3}) are equivalent. Since $S_k$ is positive-definite if and only if $k=1$, it follows immediately that $S$ is positive-definite if and only if the blocks $S_{2n_1},\cdots,S_{2n_p}$ vanish, $S_{n_q},\cdots,S_{n_r}$ are of order one, and $\epsilon_{n_q},\cdots,\epsilon_{n_r}=1$. By Theorem \ref{thmheta}, this is true only if $J$ is a real diagonal matrix ($H$ is unbroken). Thus,
 a $\cal PT$-symmetric operator $\cal H$ has a positive-definite metric operator $\eta$ if and only if $\cal H$ is unbroken.

\section{Embedding of $\cal H$}

Below we give a mathematical description of the embedding property.
Let $\cal H$ be a $\cal PT$-symmetric operator on $\mathbb C^n$ and
$\hat{\cal H}$ be a Hermitian operator on $\mathbb{C}^m$, where
$m>n$. ${\cal P}_1$ is an operator defined by
${\cal P}_1:\mathbb{C}^m\rightarrow\mathbb{C}^n$, ${\cal P}_1
\bpm\begin{smallmatrix}\phi_1\\\phi_2\end{smallmatrix}\epm=\phi_1$, where $\phi_1\in \mathbb{C}^n$ and $\phi_2\in \mathbb{C}^{m-n}$. Let
$X_{\hat{\cal H}}=\{x: x\in \mathbb C^m, {\cal P}_1\hat{\cal H}
x={\cal H} {\cal P}_1x, {\cal P}_1e^{-it\hat{\cal H}}x=e^{-it\cal
H}{\cal P}_1 x\}.$
If ${\cal P}_1X_{\hat{\cal H}}=\mathbb{C}^n$, then we say that $\cal
H$ has the embedding property and $\hat{\cal H}$ is a Hermitian
dilation of $\cal H$. We also say that $\cal H$ can be dilated to $\hat{\cal H}$.
Here, in the background of above mathematical definition, the embedding or Hermitian dilation can be seen as a special kind of map.
%
Authors in Ref. \cite{gunther2008naimark}, by utilizing the property
$\eta+\eta^{-1}=tI_2$ of a metric operator $\eta$,
dilated a class of unbroken operators on $\mathbb C^2$.
However, the following lemma tells us that this approach may not be
true in general.

\begin{lem}\label{31}
For each unbroken operator $\cal H$ on $\mathbb{C}^n$ ($n\leq 2$), there exists
a metric operator $\eta$ and a constant $t>0$ such that
$\eta+\eta^{-1}=tI$. For $n\geqslant 3$, there exists an unbroken
operator $\cal H$ such that for any metric operator $\eta$,
$\eta+\eta^{-1}\neq tI$.
\end{lem}

\begin{proof} It is clear for $n=1$. We consider the $n=2$ case.
By assumption, there exists a positive-definite operator $\eta'>0$
such that $H^\dag\eta'=\eta' H$. Hence one can take a constant $c_0$
such that $det(c_0\eta')=c_0^2det(\eta')=1$. The two eigenvalues of
$c_0\eta'$ are $\mu$ and ${1\over\mu}$. If we take $\eta=c_0\eta'$ then
$\eta+\eta^{-1}=(\mu + {1\over\mu})I_2$, and the conclusion follows.
Next, we consider the $n=3$ case. Take
$Q=\bpm\begin{smallmatrix}1&1&1\\0&1&1\\0&0&1\end{smallmatrix}\epm$,
$\Sigma=\bpm\begin{smallmatrix}1&0&0\\0&2&0\\0&0&3\end{smallmatrix}\epm$,
$H_3=Q\Sigma Q^{-1}=\bpm\begin{smallmatrix}1&1&1\\0&2&1\\0&0&3\end{smallmatrix}\epm$ and $P=T=I$. 
Apparently, $PT\overline{H}=HPT$. Hence $H_3$ is $\cal PT$-symmetric. Moreover, since $\Sigma$ is a real diagonal matrix, $H_3$ actually defines an unbroken $\cal PT$-symmetric operator.
Let $\eta$ be a positive-definite matrix satisfying $\eta
H_3=H_3^\dag \eta$. We assert that $Q^\dag\eta Q$ is a diagonal matrix (see Ref. \cite{note1}).
%
Let $Q^\dag \eta Q \equiv A=\bpm
\begin{smallmatrix}a_1&0&0\\0&a_2&0\\0&0&a_3\end{smallmatrix}\epm$,
where $a_i>0$ ($i=1, 2, 3$).
Provided $\eta+\eta^{-1}=tI_3$ holds, we have $\eta^2+I_3=t\eta$ and
$AQ^{-1}(Q^{-1})^\dag A+Q^\dag Q= tA$.
However, direct calculation shows that $ AQ^{-1}(Q^{-1})^\dag
A+Q^\dag Q=\bpm\begin{smallmatrix}
2a_1^2+1&1-a_1a_2&1\\1-a_1a_2&2a_2^2+2&2-a_2a_3\\1&2-a_2b_3&a_3^2+3\end{smallmatrix}\epm\neq
tA $, which is a contradiction.
Finally, we consider the $n>3$ case. Take $H_n=\bpm\begin{smallmatrix}
H_3&0\\0&\alpha_0I_{n-3}\end{smallmatrix}\epm$, where $\alpha_0$ is
not an eigenvalue
of $H_3^\dag$. Suppose that $\eta_n=\bpm\begin{smallmatrix}\eta_{3}&\eta_{3,n-3}\\
\eta_{3,n-3}^\dag&\eta_{n-3}\end{smallmatrix}\epm$ is a positive-definite matrix,
where $\eta_{3}$ is a matrix of order $3$ and
$\eta_{n-3}$ is a matrix of order $n-3$. $H_n^\dag\eta_n=\eta_n H_n$
yields $H_3^\dag\eta_{3,n-3}=\alpha_0\eta_{3,n-3}$. Since $\alpha_0$ is not an eigenvalue of $H_3^\dag$, we have
$\eta_{3,n-3}=0$. Now, if $\eta_n+\eta_n^{-1}=tI$ holds, we have
$\eta_{3}+\eta_{3}^{-1}=tI_3$. This contradicts with the
discussion of $H_3$ above.
\end{proof}

Although the above $\eta$-approach doesn't ensure the dilation of unbroken operators on $\mathbb C^n$, the following theorem completely characterizes the embedding property of $\cal PT$-symmetric operators.
\begin{thrm}
 A $\cal PT$-symmetric operator $\cal H$ on $\mathbb C^n$ has the embedding property if and only if $\cal H$ is unbroken.
\label{Thm41}
\end{thrm}
\begin{proof} Firstly, we consider the necessary part. Let $\hat{\cal H}$ be the Hermitian dilation of $\cal H$. We assert that
\be
X_{\hat{\cal H}}=\{x: x\in \mathbb C^m, {\cal P}_1\hat{\cal H}^k x={\cal H}^k{\cal P}_1x, k\in \mathbb N\}. \label{Hn}
\ee

By definition, for $x\in X_{\hat{\cal H}}$ we have ${\cal P}_1\sum_{k=2}^{\infty} \frac{(-it\hat{\cal H})^k}{k!} x=\sum_{k=2}^{\infty} \frac{(-it{\cal H})^k}{k!}{\cal P}_1 x$. It then follows that
\begin{eqnarray}
\nonumber&&{\cal P}_1\sum_{k=3}^{\infty} \frac{(-i\hat{\cal H})^k}{k!}t^{k-2} x-\sum_{k=3}^{\infty} \frac{(-i{\cal H})^k}{k!}t^{k-2}{\cal P}_1 x\\
&=&-\frac{ {\cal H}^2}{2}{\cal P}_1 x+{\cal P}_1 \frac{\hat{\cal H}^2}{2}x.\label{x_0}
\end{eqnarray}

Suppose that there exists a unit vector $x_0\in X_{\hat{\cal H}}$ such that ${\cal P}_1\hat{\cal H}^2 x_0\neq{\cal H}^2 {\cal P}_1 x_0$. Note that in $\mathbb C^k$ both $H$ and $\hat{H}$ are bounded operators, and one can always find a $t_0$ such that for $t<t_0$,
\begin{eqnarray*}
&&t\|\hat{H}\|<1,~~t\|H\|<1 ~~\text{and}\\
&&q_3t(\|\hat{H}\|^3+\|H\|^3)<\|-\frac{ {\cal H}^2}{2}{\cal P}_1 x_0+{\cal P}_1 \frac{\hat{\cal H}^2}{2}x_0\|,
\end{eqnarray*}
where $q_3=\displaystyle\sum_{k=3}^\infty (1/k!)$.
Now, using the triangle inequality, one can obtain
$\|({\cal P}_1\sum_{k=3}^{\infty} \frac{(-i\hat{\cal H})^k}{k!}t^{k-2} x_0-\sum_{k=3}^{\infty} \frac{(-i{\cal H})^k}{k!}t^{k-2}{\cal P}_1 x_0)\|
\leqslant q_3t(\|\hat{H}\|^3+\|H\|^3)
<\|-\frac{ {\cal H}^2}{2}{\cal P}_1 x_0+{\cal P}_1 \frac{\hat{\cal H}^2}{2}x_0\|$,
which contradicts with Eq. (\ref{x_0}). This shows that ${\cal P}_1\hat{\cal H}^2 x={\cal H}^2 {\cal P}_1 x$.
For $k>2$, this can be proved by induction. Thus, we have $X_{\hat{\cal H}}\subseteq \{x: x\in \mathbb C^m, {\cal P}_1\hat{\cal H}^k x={\cal H}^k{\cal P}_1x, k\in \mathbb N\}$. The converse containing relation is clear, which validates Eq. (\ref{Hn}).
Now, if $x\in X_{\hat{\cal H}}$ and
$y=\hat{\cal H}x$ then ${\cal P}_1\hat{\cal H}^k y={\cal H}^{k+1}{\cal P}_1 x={\cal H}^k
{\cal P}_1 y$. Hence $y\in X_{\hat{\cal H}}$. Thus $\hat{\cal H}X_{\hat{\cal H}}\subseteq
X_{\hat{\cal H}}$. Since $\hat{\cal H}$ is a Hermitian operator on $\mathbb C^m$, $\hat{\cal H}|X_{\hat{\cal H}}$ is also a Hermitian operator on $X_{\hat{\cal H}}$.
Let $X_0=Ker({\cal P}_1|X_{\hat{\cal H}})$ and $X_0^{\bot}$ be the orthogonal complement of $X_0$ in $X_{\hat{\cal H}}$. It is apparent that $X_0$ is an invariant
subspace of $\hat{\cal H}$. Since $\hat{\cal H}|X_{\hat{\cal H}}$ is Hermitian
and $X_0$ is an invariant subspace of $\hat{\cal H}|X_{\hat{\cal H}}$, this implies that $X_0^{\bot}$
is also an invariant subspace of $\hat{\cal H}|X_{\hat{\cal H}}$. Thus, there is a basis $\{u_i\}_{i=1}^l$ of $X_0^{\bot}$ such that $\hat{\cal H}u_i=a_iu_i$, where $a_i$'s are real numbers. By assumption, ${\cal H}({\cal P}_1u_i)={\cal P}_1(\hat{\cal
H}u_i)=a_i{\cal P}_1u_i$, which implies that $\cal H$ is unbroken. Hence, the necessary part is proved.

Now, we prove the sufficient part. Let $\cal H$ be an unbroken operator on $\mathbb C^n$. Then it is enough to show that there exists a Hermitian operator $\hat{\cal H}$
on $\mathbb C^{2n}$ and an $n$ dimensional subspace $Y$ of $\mathbb C^{2n}$ such that $Y\subseteq X_{\hat{\cal H}}$ and ${\cal P}_1 Y=\mathbb C^n$.
First note that if an $n$ dimensional subspace $Z$ of $\mathbb C^{2n}$ satisfies the condition ${\cal P}_1(Z)=\mathbb C^{n}$, then there exists a unique $n\times n$ matrix $\tau$ such that $Z=Y_{\tau}=\{\hat{\psi}_\tau:\hat{\psi}_\tau=\begin{pmatrix}\begin{smallmatrix}\psi\\
\tau\psi\end{smallmatrix}\end{pmatrix}, \psi\in \mathbb C^n\}$.
Thus, the problem reduces to finding a $\tau$ and an $\hat{H}$ such that $Y_\tau\subseteq X_{\hat{\cal H}}$.
On the other hand, a Hermitian matrix $\hat{H}$ of order $2n$ has the form $\hat{H}=\bpm \begin{smallmatrix}H_1&H_2\\H_2^\dag& H_4 \end{smallmatrix}\epm$, where $H_1$ and $H_4$ are $n\times n$ Hermitian matrices. Let $H$ be the matrix of $\cal H$. It can be shown that $Y_{\tau}\subseteq X_{\hat{H}}$
if and only if the following conditions are simultaneously satisfied
\begin{numcases}{}
   H_1+H_2\tau=H,\label{H1H2} \\
   H_2^\dag+H_4\tau=\tau H.\label{H2H4}
\end{numcases}

Since $\cal H$ is unbroken, there is a positive-definite matrix $\eta'$
such that $H^\dag\eta'=\eta' H$. We choose a positive number $t$ such
that $\eta=t\eta'>I$. Now let $H_1$ be an arbitrary $n\times n$ Hermitian matrix and
\begin{eqnarray}
&&\tau=(\eta-I)^{1\over 2},\label{sigma}\\
&&H_2=(H-H_1)\tau^{-1},\label{H2}\\
&&H_4=(\tau H-H_2^\dag)\tau^{-1}\label{H4}.
\end{eqnarray}

It can be seen that $H_i$ and $\tau$ satisfy Eqs. (\ref{H1H2}) and (\ref{H2H4}).
Thus $\hat{H}=\begin{pmatrix}\begin{smallmatrix}
H_1&H_2\\
H_2^\dag&H_4\end{smallmatrix}
\end{pmatrix}$ is a Hermitian dilation of $H$ and $Y_{\tau}\subseteq X_{\hat{\H}}$. This completes the proof.
\end{proof}

Eqs. (\ref{H1H2}) and (\ref{H2H4})
completely characterize the embedding property. For a state $\hat{\psi}_\tau=\bpm\psi\\
\tau\psi\epm\in Y_\tau\subseteq X_{\hat{\H}}$,
\be
\hat{U}(t)\hat{\psi}_\tau
=
\begin{smallmatrix}\bpm U(t)&0\\0&\tau U(t)\tau^{-1}\epm\bpm\psi\\\tau\psi\epm
=\bpm U(t)\psi\\ \tau U(t)\psi\epm.\end{smallmatrix}
\label{U}
\ee

Note that $ \hat{U}(t)\hat{\psi}_\tau=\bpm \begin{smallmatrix}U(t)\psi\\ \tau U(t)\psi\end{smallmatrix}\epm$ has a component $U(t)\psi$ in $\mathbb C^n$. Thus, the evolution $ U(t)$ in $\mathbb C^n$ is realized by  the Hermitian Hamiltonian $\hat{H}$
on $\mathbb C^{2n}$ and the vectors in $X_{\hat{H}}$.

\section{The simulation of $\cal H$ by utilizing the Hermitian dilation $\hat{{\cal H}}$}

\subsection{The simulation paradigm}
In this section we use Theorem \ref{Thm41} to devise a simulation paradigm, which gives a general method to experimentally realize an arbitrary unbroken $\cal PT$-symmetric Hamiltonian in a finite dimensional space.
Put another way, we specify how a $\cal PT$-symmetric Hamiltonian $\cal H$ on $\mathbb C^{n}$ can experimentally be connected to a Hermitian dilation $\hat{\cal H}$ on a large space $\mathbb C^{2n}$.
To this end, we need the following lemma. This lemma can be derived from the Naimark dilation theorem (see
Ref. \cite{croke2015pt}). However, to show how the quantum states can be prepared, we prove it in detail.

\begin{lem}\cite{croke2015pt}\label{lem6} Let $M$ and $N$ be two $n$ dimensional subspaces of $\mathbb C^{2n}$ and
${\cal A}:M\rightarrow N$ be a linear operator. Define a new
transformation $\tilde{{\cal A}}:M\rightarrow N$ by
\begin{equation}
\tilde{{\cal A}}\ket{\nu}=\left\{\begin{aligned}&\frac{1}{\|{\cal A}\ket{\nu}\|}{\cal A}\ket{\nu},&& \text{if }{\cal A}\ket{\nu}\neq 0,\\
&0, &&\text{if }{\cal A}\ket{\nu}=0,\end{aligned}\right. \label{61}
\end{equation}
Then $\tilde{{\cal A}}$ can be realized by a unitary operator and an orthogonal projection operator.
\end{lem}

\begin{proof}
Let $N^\perp$ be the orthogonal complement of $N$ in $\mathbb C^{2n}$ and $P_N$ be the orthogonal projection operator onto $N$.
If ${\cal A}=0$, since $N$ and $N^\perp$ have the same dimension, we can take a unitary operator
${\cal U}_{\cal A}$ such that
${\cal U}_{\cal A}(N)=N^\perp$. It is apparent that $P_N{\cal U}_{\cal A}\ket{\nu}=\tilde{{\cal A}}\ket{\nu}=0$.
If ${\cal A}\neq 0$, we take an orthonormal basis $\{\ket{u_i}\}_{i=1}^n$ of $M$ and
denote $\ket{g_i}=\frac{{\cal A}\ket{u_i}}{\sqrt{\sum_j \|{\cal A}\ket{u_j}\|^2}}$.
Note that we can construct $\{\ket{h_i}\}_{i=1}^n$ such that $\braket{g_i+h_i|g_j+h_j}=\delta_{ij}$.
In fact, let $\{\ket{v_i}\}_{i=1}^n$ be an orthonormal basis of $N$ and $\{\ket{w_i}\}_{i=1}^n$ be
an orthonormal basis of $N^\perp$. Then $\ket{g_j}=\sum_i c_{ij}\ket{v_i}$, where
$C=(c_{ij})_{n\times n}$ is a matrix. By the property of the operator norm $\|\cdot\|_\infty$,  $\|C^\dag C\|_\infty\leqslant tr (C^\dag C)=\displaystyle\sum_i \|\ket{g_i}\|^2=1$. Hence $C^\dag C\leqslant I_k$. We choose a matrix $D=(d_{ij})_{n\times n}$ such that $C^\dag C+D^\dag D=I_k$.
For $\ket{h_j}=\sum_i d_{ij}\ket{w_i}$, it is easy to verify that $\braket{g_i+h_i|g_j+h_j}=\delta_{ij}$.
Thus, one can define a unitary operator ${\cal U}_{\cal A}$ such that ${\cal U}_{\cal A}\ket{u_i}= \ket{g_i}+ \ket{h_i}$.
 Let $\ket{\nu}=\displaystyle\sum_i a_i\ket{u_i}\in M$.
Then ${\cal U}_{\cal A}\ket{\nu}=
\sum_i a_i (\ket{g_i}+ \ket{h_i})$. By projecting ${\cal U}_{\cal A}\ket{\nu}$ onto $N$ and normalizing the vector, we can obtain (\ref{61}). For convenience, we will assume $\frac{1}{\|{\cal A}\ket{\nu}\|}{\cal A}\ket{\nu}=0$ if ${\cal A}\ket{\nu}=0$.
\end{proof}

\begin{remark} To experimentally realize the non-unitary transformation in Eq. (\ref{61}), one can first prepare a Hermitian Hamiltonian $\cal H_{A}$ on $\mathbb C^{2n}$ to generalize $\cal U_{A}$. The evolution of $\ket{\nu}$ under $\cal H_{A}$ will give a result state ${\cal U}_A\ket{\nu}$. Recall that $N$ and
$N^\perp$ are orthogonal subspaces of ${\mathbb C}^{2n}$. The corresponding projections $P_N$ and $P_N^\perp$ form a von Neumann measurement on ${\mathbb C}^{2n}$, which can be employed to measure the state ${\cal U}_A\ket{\nu}$. Moreover, by post-selecting the measured state in $N$, one can experimentally obtain the final state in Eq. (\ref{61}) (see also \cite{note2}).\end{remark}

%

The proposed simulation paradigm has three stages: (1) the pre-simulation stage, (2) the simulation stage and (3) the post-simulation stage. Let $\ket{\xi}$ be a state in $\mathbb C^n$.
(1) The pre-simulation stage is to prepare a state that will be acted upon by the Hermitian dilation Hamiltonian $\hat{H}$ later in the simulation stage. In this stage, the state $\ket{\xi}$ is transferred to a state in subspace $Y_{\tau}$ in the following two steps. (i) The state $\ket{\xi}$ is transferred to a state $\ket{\xi^{(1)}}$ in $\mathbb C^{2n}$ by coupling the system under investigation to an ancillary system in $\mathbb C^{2}$,
\begin{equation}
\ket{\xi}\mapsto\ket{\xi^{(1)}}=\bpm \ket{\xi}\\ {0}\epm=\ket{0}\ket{\xi},\label{62}
\end{equation}
where $\ket{0}=\begin{smallmatrix}\bpm 1\\0\epm\end{smallmatrix}$ and $0$ is the zero vector of $\mathbb C^n$.
(ii) The state $\ket{\xi^{(1)}}$ is transferred to a state $\ket{\xi^{(2)}}$ in the space $Y_{\tau}$,
\begin{equation}
\ket{\xi^{(1)}}\mapsto\ket{\xi^{(2)}}=\frac{1}{\|\sqrt{\eta}\rho\ket{\xi}\|}\bpm \rho \ket{\xi}\\ \tau \rho\ket{\xi}\epm,\label{63}
\end{equation}
where $\rho$ is a matrix determined by the environment or experimental conditions.
In appropriate setting of the experiment, $\rho$ can be any matrix.
To see this, let $X_1=\{\bpm\begin{smallmatrix} \ket{\phi}\\ {0}\end{smallmatrix}\epm:\ket{\phi}\in \mathbb C^n\}$. Note that any given matrix $\rho$ can induce a linear operator $\hat{\rho}$:
$\bpm \begin{smallmatrix}\ket{\xi}\\ {0}\end{smallmatrix}\epm \in X_1 \mapsto\bpm \begin{smallmatrix}\rho \ket{\xi}\\ \tau \rho\ket{\xi}\end{smallmatrix}\epm \in Y_{\tau}$.
Hence Eq. (\ref{63}) is actually $\ket{\xi}\mapsto \frac{1}{\|{\hat{\rho}}\ket{\xi}\|}{\hat{\rho}}\ket{\xi}$.
By Lemma \ref{lem6}, it is realized by a unitary operator
${\cal U}_\rho$ and the orthogonal projection onto the space $Y_\tau$. As was discussed in the remark after Lemma \ref{lem6}, this procedure can be realized in experiments.

(2) In the simulation stage, the evolution of the Hermitian dilation Hamiltonian $\hat{H}$ on $\mathbb C^{2n}$ visualises the evolution of the $\cal PT$-symmetric Hamiltonian ( $U(t)=e^{-itH}$ ) in the subspace $\mathbb C^n$.
In this stage, $\ket{\xi^{(2)}}$, the state prepared in the pre-simulation stage is acted upon by the Hermitian dilation Hamiltonian $\hat{H}$. Thus, $\ket{\xi^{(2)}}$ is transformed to  $\ket{\xi^{(3)}}=\hat{U}(t)\ket{\xi^{(2)}}$, where $\hat{U}(t)=e^{-it\hat{H}}$.
From Eq. (\ref{U}),
\begin{eqnarray}
&&\ket{\xi^{(2)}}\mapsto\ket{\xi^{(3)}}=\hat{U}(t)\ket{\xi^{(2)}}\notag\\
&&=\frac{1}{\|\sqrt{\eta}U(t)\rho\ket{\xi}\|}\bpm U(t)\rho \ket{\xi}\\ \tau U(t)\rho\ket{\xi}\epm.\label{64}
\end{eqnarray}
Since $\hat{H}$ is Hermitian, it is apparent that such a process is realizable in practice.

(3) The post-simulation stage retains our original system ($\mathbb C^n$) and the effect of $\cal PT$-symmetric Hamiltonian $H$ (the evolution $U(t)$) in it, by removing the ancillary system.
It consists of the following two steps in which the state $\ket{\xi^{(3)}}$ is transformed to a state $\ket{\xi^{(5)}}$ in $\mathbb C^n$.
(i) The state $\ket{\xi^{(3)}}$ evolves to a state $\ket{\xi^{(4)}}$ in $\mathbb C^{2n}$,
\begin{eqnarray}
&&\ket{\xi^{(3)}}\mapsto\ket{\xi^{(4)}}\notag\\
&&=\frac{1}{\|\rho'U(t)\rho\ket{\xi}\|}\bpm \rho'U(t)\rho \ket{\xi}\\ 0\epm,\label{65}
\end{eqnarray}
where $\rho'$ is also determined by the environment or experimental conditions. Note that $\rho'$ induces a linear operator $\hat{\rho}'$:
$\ket{\xi^{(3)}} \in Y_\tau \mapsto \bpm \rho'U(t)\rho \ket{\xi}\\ 0\epm \in X_1$. Eq. (\ref{65}) is actually
$\ket{\xi^{(3)}}\mapsto \frac{1}{\|\hat{\rho}'\ket{\xi^{(3)}}\|}\hat{\rho}'\ket{\xi^{(3)}}$.
By Lemma \ref{lem6}, this can be simulated in the experiment.
(ii) The state $\ket{\xi^{(4)}}$ is transformed to the final state $\ket{\xi^{(5)}}$ in $\mathbb C^n$,
\begin{equation}
\ket{\xi^{(4)}}\mapsto\ket{\xi^{(5)}}=\frac{1}{\|\rho'U(t)\rho\ket{\xi}\|} \rho'U(t)\rho \ket{\xi}.\label{66}
\end{equation}

In fact, $\ket{\xi^{(4)}}=\ket{0}\ket{\xi^{(5)}}$. By making a von Neumann measurement $\{\ket{0}\bra{0}, \ket{1}\bra{1}\}$, post-selecting the local state $\ket{0}$ on the ancillary system, and finally removing it,
 one can obtain the state $\ket{\xi^{(5)}}$. Thus, in the simulation process, following Eqs. (\ref{62})-(\ref{66}) the final state is given by
\be
\ket{\xi}\mapsto \frac{\rho'U(t)\rho\ket{\xi}}{\|\rho'U(t)\rho\ket{\xi}\|}.\label{67}
\ee

One can observe that in the simulation paradigm presented above, not only the effect of the $\cal PT$-symmetric Hamiltonian $U(t)$ but also the effects of $\rho$ and
$\rho'$ are impressed in the system. The introduction of the two operators $\rho$ and $\rho'$ in our simulation paradigm is motivated by Croke \cite{croke2015pt}. Intuitively, they represent certain interactions before and after the simulation stage. In particular, if we take $\rho=\eta^{-\frac{1}{2}}$ and $\rho'=\eta^{\frac{1}{2}}$, the state will transform to $\ket{\xi}\mapsto \eta^{\frac{1}{2}}U(t)\eta^{-\frac{1}{2}}\ket{\xi}$. This experimentally realizes the change of the framework by a similarity transformation.
And, if we take $\rho'=\rho=I$, the state will transform to $\ket{\xi}\mapsto U(t)\ket{\xi}$. In this case, the effect of the simulation is completely given by the $\cal PT$-symmetric evolution.
It should be noted that in the above simulation paradigm, both the pre-simulation and the post-simulation stages are non-unitary. Hence the simulated $\cal PT$-symmetric system is actually open. That is, one can realize the simulation only probabilistically in practice.

\begin{remark}
Hermitian dilation is a key to the simulation paradigm. First by transferring the state of system of interest in $\mathbb C^{n}$ to one in $\mathbb C^{2n}$ by coupling it with an ancillary system, then evolving it with a Hermitian Hamiltonian $\hat{\cal H}$ on $\mathbb C^{2n}$, and finally decoupling the ancillary system, we address the problem of how to naturally realize a simulation of $\cal PT$-symmetric systems in the subspace $\mathbb C^n$. We emphasize that there exists essential difference between a similarity transform (or the more general case $\rho'U(t)\rho$) and the embedding property. The embedding is a procedure of introducing a Hermitian Hamiltonian $\hat{\cal H}$ in a large space $\mathbb C^{2n}$ with essential constraints in the form of Eqs. (\ref{H1H2}), (\ref{H2H4}) and (\ref{U}), while the former concerns only with the subspace $\mathbb C^n$. It is one possible result of the simulation paradigm (by removing the ancillary system). That is, the widely known similarity transform $\eta^{\frac{1}{2}}H\eta^{-\frac{1}{2}}$ (or $\eta^{\frac{1}{2}}U(t)\eta^{-\frac{1}{2}}$) is not a synonym for embedding. For the same reason, Eq. (\ref{67}) is itself not embedding but a result of the simulation paradigm.
\end{remark}

\subsection{An example}
In this subsection, we illustrate that the simulation paradigm outlined in subsection IV A can be transferred to a physical system. To give an example, we consider the system in Ref. \cite{gunther2008naimark} and ``map'' it to the simulation paradigm. 
As mentioned above, many procedures are non-unitary. Hence, for convenience, we often do not normalize the state following Ref. \cite{gunther2008naimark}. 

Let us recall that in Ref. \cite{gunther2008naimark}, the target was to simulate the effect of a $\cal PT$-symmetric Hamiltonian $H=\bpm\begin{smallmatrix}
E_0+is\sin\alpha&s\\s&E_0-is\sin\alpha\end{smallmatrix}\epm$ on the state $\ket{\psi_I}=\begin{smallmatrix}\bpm 1\\0\epm\end{smallmatrix}$. To this end, a concrete Hermitian dilation operator $\hat{H}$ and its unitary evolution operator $\hat{U}(t)$ on $\mathbb C^4$ were figured out at first. Then, by intricately choosing the initial state $\ket{\hat{\psi}_I}=\begin{smallmatrix}\bpm\ket{\psi_I}\\ \ket{\chi_I}\epm\end{smallmatrix}$, where $\ket{\chi_I}=\frac{1}{\cos\alpha}\begin{smallmatrix}\bpm 1\\i\sin\alpha\epm\end{smallmatrix}$, one can obtain a final state  $\hat{U}(t)\ket{\hat{\psi}_I}= \begin{smallmatrix}\bpm \ket{\psi(t)}\\ \ket{\chi(t)}\epm\end{smallmatrix}$. Moreover, $\ket{\psi(t)}=e^{-itH}\ket{\psi_I}$. Thus, the  simulation of a $\cal PT$-symmetric Hamiltonian was visualised in the subspace.

To imitate the above discussion in the simulation paradigm, there are two key points to note. First, the Hermitian Hamiltonian $\hat{H}$ on $\mathbb C^4$ in Ref. \cite{gunther2008naimark} is a concrete example of the construction of Hermitian dilation operators in Theorem \ref{thmhpt}.
To see this, take
$\eta=\frac{2}{\cos^2\alpha}\bpm\begin{smallmatrix}
1&-i\sin\alpha\\i\sin\alpha &1\end{smallmatrix}\epm$,
$H=\bpm\begin{smallmatrix}
E_0+is\sin\alpha&s\\s&E_0-is\sin\alpha\end{smallmatrix}\epm$,
$H_1=\tau H\tau\eta^{-1}+H\eta^{-1}$ in Eqs.
(\ref{sigma})-(\ref{H4}). After some tedious calculations, we can obtain

$\tau=\frac{\cos\alpha}{2}\eta=\frac{1}{\cos\alpha}\begin{smallmatrix}\bpm 1&-i\sin\alpha\\i\sin\alpha& 1\epm\end{smallmatrix}$,

$\begin{smallmatrix}H_1=\frac{\cos\alpha}{2}(\tau H+H\tau^{-1})=\bpm E_0& s\cos^2\alpha\\ s\cos^2\alpha &E_0\epm\end{smallmatrix}$,

$\begin{smallmatrix}H_2=(H-H_1)\tau^{-1}=
\bpm
is\sin\alpha\cos\alpha&0\\
0 & -is\sin\alpha\cos\alpha
\epm\end{smallmatrix}$,

$\begin{smallmatrix}H_4=(\tau H-H_2^\dag)\tau^{-1}=
\bpm
E_0&s\cos^2\alpha\\
s\cos^2\alpha& E_0
\epm
\end{smallmatrix}
$.

Then the Hamiltonian $\hat{H}=\bpm H_1& H_2\\H_2^\dag& H_4\epm$ can be rewritten as $\hat{H}=I_2\otimes (E_0 I_2+s\cos^2\alpha\sigma_x)-\cos\alpha\sin\alpha\sigma_y\otimes\sigma_z$, which was
given in Ref. \cite{gunther2008naimark}.

Second, the pre-simulation stage can be utilized to obtain the intricately chosen state $\ket{\hat{\psi}_I}=\bpm\ket{\psi_I}\\ \ket{\chi_I}\epm$ on $\mathbb C^4$.
Firstly, by preparing the state $\ket{\psi_I}$ on $\mathbb C^2$ and coupling the system to an ancillary system $\mathbb C^2$ (e.g. a $2$-spin system), one can obtain a state $\bpm\ket{\psi_I}\\0\epm$.
Note that $\tau(\ket{\psi_I})=\ket{\chi_I}$ and $\bpm\ket{\psi_I}\\ \ket{\chi_I}\epm$ is actually a state in $Y_\tau$.
Hence the state transformation  $\begin{smallmatrix}\bpm\ket{\psi_I}\\0\epm\mapsto \bpm\ket{\psi_I}\\ \ket{\chi_I}\epm\end{smallmatrix}$
can be viewed as a special case of Eq. (\ref{63}), by taking $\rho=I$ and $\ket{\nu}=\ket{\psi_I}$. Thus, the chosen initial state $\ket{\hat{\psi_I}}$ in \cite{gunther2008naimark} can be obtained in the pre-simulation stage. 
The prescription is as follows.

Take a unitary matrix
\be
U_\tau=\frac{1}{2}
\begin{smallmatrix}
\bpm
e^{i\alpha}&-1&1&e^{i\alpha}\\
-1&e^{-i\alpha}&e^{-i\alpha}&1\\
1&e^{-i\alpha}&-e^{-i\alpha}&1\\
e^{i\alpha}&1&1&-e^{i\alpha}
\epm.
\end{smallmatrix}
\label{U_tau}
\ee

Note that Stone theorem \cite{reedsimon1980methods} ensures that $U_\tau$ is the evolution operator corresponding to some Hermitian Hamiltonian $H_\tau$ at some time $t_0$.
Such a Hermitian Hamiltonian can be appropriately constructed.
Let the state $\begin{smallmatrix}\bpm\ket{\psi_I}\\ 0\epm\end{smallmatrix}$ be evolved under $H_\tau$ to give $U_\tau \begin{smallmatrix} \bpm\ket{\psi_I}\\ 0\epm\end{smallmatrix}$.

Consider two orthogonal projection operators
\be
P_{Y_\tau}=
\begin{smallmatrix}
\frac{1}{2}\bpm
1&i\sin\alpha&\cos\alpha&0\\
-i\sin\alpha& 1&0&\cos\alpha\\
\cos\alpha&0&1&-i\sin\alpha\\
0&\cos\alpha&i\sin\alpha& 1
\epm
\end{smallmatrix},
\ee

and 

\be
P_{Y_\tau}^\perp=
\begin{smallmatrix}
\frac{1}{2}\bpm
1&-i\sin\alpha&-\cos\alpha&0\\
i\sin\alpha& 1&0&-\cos\alpha\\
-\cos\alpha&0&1&i\sin\alpha\\
0&-\cos\alpha&-i\sin\alpha& 1
\epm
\end{smallmatrix}.
\ee

It can be verified that $P_{Y_\tau}$ is the projection onto subspace $Y_\tau$ and $P_{Y_\tau}^\perp+P_{Y_\tau}=I$. Hence, one can conduct a von Neumann measurement $\{P_{Y_\tau}, P_{Y_\tau}^\perp\}$ on the state $U_\tau \begin{smallmatrix} \bpm\ket{\psi_I}\\ 0\epm\end{smallmatrix}$
and post-select the resulting state in $Y_\tau$. Direct calculation shows that
 \be
  P_{Y_\tau}U_\tau \begin{smallmatrix} \bpm\ket{\psi_I}\\ 0\epm\end{smallmatrix}=\frac{\cos\alpha}{2} \ket{\hat{\psi}_I}.
  \label{effi}
 \ee
 In this way, the state $\ket{\hat{\psi}_I}$ is obtained in the pre-simulation stage.

Now that we have prepared the chosen initial state $\ket{\hat{\psi_I}}$ in \cite{gunther2008naimark}, we can further apply the Hermitian Hamiltonian $\hat{H}$ and the simulation paradigm to this state. The effect of such an evolution is given by Eq. (\ref{U}).
 Since we utilize the same Hamiltonian $\hat{H}$ and state $\ket{\hat{\psi_I}}$, we will obtain the same result as in \cite{gunther2008naimark}.
After the simulation stage, the state will be $\bpm U(t)\ket{\psi_I}\\ \tau U(t)\ket{\psi_I}\epm=\ket{0}\otimes U(t)\ket{\psi_I}+\ket{1}\otimes \tau U(t)\ket{\psi_I}$. 
We can then conduct a measurement $\{\ket{0}\bra{0},\ket{1}\bra{1}\}$ on the ancillary system, post-select it in state $\ket{0}$ and finally discard it. The post-simulation stage produces the state $U(t)\ket{\psi_I}$ and helps see
the effect of a $\cal PT$-symmetric Hamiltonian in $\mathbb C^2$, which was also proposed in \cite{gunther2008naimark}.

It should be noted that the simulation processes in this subsection concern with spaces $\mathbb C^2$ and $\mathbb C^2 \otimes \mathbb C^2$ which are often relatively easier to be devised by utilizing optical systems. In this case, $\tau$, $\rho$, $\rho'$ and the Hermitian dilation $\hat{H}$ can all be elaborated by the arrangement of the optical instruments.
In particular, the key procedure to post-select the state in $\ket{\hat{\psi_I}}$ and fulfill the pre-simulation stage, can be realized by using beam splitters. In that context, $\ket{\hat{\psi_I}}$ and the state orthogonal to it correspond to two different beams of light. The beam splitter can reserve the state $\ket{\hat{\psi_I}}$ by discarding the undesired beam of light to the surroundings. For more details, see Ref. \cite{tang2016experimental}.

\begin{remark}
The concrete example of $U_\tau$ in Eq. (\ref{U_tau}) is actually obtained by using the method described in the proof of Lemma \ref{lem6},
while the tedious calculations are omitted here. However, in general, there exist other ways to construct different unitary matrices $U_\tau$ from that in Eq. (\ref{U_tau}).
On the other hand, there is a constant $\frac{\cos\alpha}{2}$ before $\ket{\hat{\psi_I}}$. Since the preparation procedure of $\ket{\hat{\psi_I}}$ is
non-unitary, $\frac{\cos\alpha}{2}$ apparently reflects the preparation efficiency. Such a constant preceding $\ket{\hat{\psi_I}}$ is determined by
$U_\tau$.  Now different choices of $U_\tau$ ( or the different arrangement of devices in experiments), will generally give different preparation efficiencies.
\end{remark}

\section{Consistency of $\cal PT$-symmetry theory}
The simulation paradigm provides a general method to investigate $\cal PT$-symmetric quantum systems by utilizing the notions and techniques in standard quantum mechanics. In this section, we further explain how the quantum processes in this paradigm can be applied to analyse the effect of the $\cal PT$-symmetric Hamiltonians. The demonstration will mainly focus on the no-signaling principle.

\subsection{The no-signaling principle}
Suppose Alice and Bob are space-like separated, and
share an entangled state on which both of them can make
measurements. Let $A_1$ and $A_2$ be two local measurements done by
Alice whose outcomes are denoted by $\{a_i^{(1)}\}_i$ and
$\{a_i^{(2)}\}_i$ respectively. $B$ is the local measurement done
by Bob whose outcomes are denoted by $\{b_j\}_j$. For each pair of
outcomes $(a_i^{(k)},b_j)$, there is a joint
probability $P(a_i^{(k)},b_j|A_k,B)$ ($k=1,2$).
From Einstein's relativity theory, we know that the no-signaling principle forbids
superluminal communication. That is, no physical information can propagate with a speed greater than that of light. Therefore, the probability distribution of Bob's measurement outcomes
must not be affected by the local operations on Alice's subsystem since they are
space-like separated. It then follows that
\begin{equation}
\displaystyle\sum_i
P(a_i^{(1)},b_j|A_1,B)=\displaystyle\sum_i
P(a_i^{(2)},b_j|A_2,B).\label{sig}\end{equation}
Eq. (\ref{sig}) is the mathematical statement of the no-signaling principle.

Authors in Ref. \cite{lee2014local} pointed out that the
$\cal PT$-symmetry theory might contradict with the no-signaling
principle. The argument is as follows.
Suppose Alice and Bob are space-like separated, both of them have a local system $\mathbb C^2$ and share an
entangled state $\ket{\psi}=\frac{1}{\sqrt{2}}(\ket{+_x +_x}+\ket{-_x -_x})$, where $\ket{\pm_x}$ are the eigenstates of
Pauli matrix $\sigma_x$. At first, Alice performs two unitary operations $U_{A_1}=I$ and $U_{A_2}=\sigma_x$ on her system. The joint states, corresponding to evolutions $U_{A_{1,2}}\otimes I_2$, are given by
$\ket{\psi^{1,2}}=[U_{A_{1,2}}\otimes I]\ket{\psi}$.
Next, Alice's system is subjected to a local unbroken $\cal PT$-symmetric Hamiltonian $H_0=s\bpm\begin{smallmatrix} i\sin\alpha&1\\1&-i\sin\alpha\end{smallmatrix}\epm$ $\left(|\alpha| < \frac{\pi}{2}\right)$. The total Hamiltonian $H_0\otimes I$ gives an evolution $[U_0(t)\otimes e^{-iIt}]\ket{\psi^{1,2}}$, where $U_0(t)=e^{-itH_0}$. The final joint states (unnormalised) are
\begin{equation}
\ket{\psi_f^{1,2}}=[U_0(t)U_{A_{1,2}}\otimes e^{-iIt}I]\ket{\psi}.\label{71}
\end{equation}

Now both Alice and Bob make local measurements on the final states with conventional projectors $\ket{\pm_y}\bra{\pm_y}$, where $\ket{\pm_y}$ are the eigenstates of
Pauli matrix $\sigma_y$.
Consequently, the no-signaling condition, Eq. (\ref{sig}) reduces to
 \be \sum_{a=\pm_{y}} P(a,+_y|A_1,B)=\sum_{a=\pm_{y}} P(a,+_y|A_2,B)\label{sig2}.\ee

But, direct calculation shows that Eq. (\ref{sig2})
does not hold, which implies a violation of the no-signaling principle \cite{lee2014local}.
%
However, there were dissenting opinions on this undesirable conclusion.
When Croke's or Brody's paradigm is applied to the above discussion, one can verify that Eq. (\ref{sig2})
is still valid. Hence, the no-signaling principle is respected by the $\cal PT$-symmetry theory \cite{croke2015pt,brody2016consistency}.

\subsection{The no-signaling principle from the perspective of simulation}
We examine the validity of the no-signaling condition, Eq. (\ref{sig2}), in the framework of our simulation paradigm.
Note that in section IV, for a single system, the $\cal PT$-symmetric Hamiltonian $H_0$ is simulated by its Hermitian dilation operator $\hat{H}_0$. Now, for a composite system, $H_0\otimes I$ will be simulated by its dilated Hermitian Hamiltonian $\hat{H}_0\otimes I$ in the simulation paradigm. Since $\hat{H}_0\otimes I$ is in a local form, one can transplant the discussions in section IV to the current situation, simply by coupling one more system $ \mathbb C^2$. Mathematically, it means that we use operators such as $U_\tau\otimes I$, $P_{Y_\tau}\otimes I$ and $P_{Y_\tau}^\perp \otimes I$ for preparing the state, and $\hat{H}_0\otimes I$ for simulation. Moreover, since both of Alice and Bob can only make local operations, it is apparent that the effect on Alice's subsystem is the same as in section IV.
Similar to Eq. (\ref{67}), the final states (after simulation) are
 \begin{equation}
\ket{\xi_f^{1,2}}=\frac{[\rho'U_0(t)\rho U_{A_{1,2}}\otimes e^{-iIt}I]\ket{\psi}}{\|[\rho'U_0(t)\rho U_{A_{1,2}}\otimes e^{-iIt}I]\ket{\psi}\|}.\label{73}
\end{equation}
Alice and Bob make local measurements on these states to obtain the probabilities, and analyze Eq. (\ref{sig2}).
It is apparent that the operators $\rho$ and $\rho'$ affect the numerical results.
If we take $\rho=\eta^{-\frac{1}{2}}$ and $\rho'=\eta^{\frac{1}{2}}$, then
the unitarity of $\eta^{\frac{1}{2}}U_0(t)\eta^{-\frac{1}{2}}$ guarantees that Eq. (\ref{sig2})
will hold. Such a scheme was also proposed in \cite{croke2015pt}.
If $\rho$ and $\rho'$ assume other forms, then $\rho'U_0(t)\rho$ is not a unitary operator in general. In particular, if $\rho'=\rho=I$, then Eq. (\ref{73}) reduces to Eq. (\ref{71}).
One can obtain an inequality $\displaystyle \sum_{a=\pm_{y}} P(a,+_y|A_1,B)\neq\displaystyle \sum_{a=\pm_{y}} P(a,+_y|A_2,B)$,
which has been verified in a recent experiment \cite{tang2016experimental}.
Thus, the simulation paradigm encompasses the existing numerical results in the literature as special cases.

However, it should be noted that breach of Eqs. (\ref{sig}) and (\ref{sig2}) in the simulation process is not essentially related to the no-signaling principle.
In fact, this is a consequence of the loss of probabilities.
For instance, similar to the example of section IV B, to obtain a final state in Eq. (\ref{73}) we need to conduct a measurement $\{P_{Y_\tau}\otimes I, P_{Y_\tau}^\perp\otimes I\}$ which will give two states, one in subspace $Y_\tau \otimes \mathbb C^2$ and another in $Y_\tau^\perp\otimes \mathbb C^2$. The state in $Y_\tau\otimes \mathbb C^2$ will be subject to $\hat{H}_0\otimes I$ for a successful simulation, giving the final state in Eq. (\ref{73}). While the state in $Y_\tau^\perp\otimes \mathbb C^2$, which is not used for simulation, is discarded by post-selection in the pre-simulation stage. This causes a loss of probabilities.
For the no-signaling principle to be observed, all the results of
all the local measurements on Alice's subsystem should be considered. However, in the simulation paradigm,
we solely consider the simulated final states (given by the prepared state only in $Y_\tau$) and thus neglect some probabilities. (Similar to the pre-simulation stage, the post-simulation stage also has such a problem.) Hence, the violation of Eqs. (\ref{sig}) and (\ref{sig2}) for the $\cal PT$-symmetric quantum systems is expected to be observed. This marks a departure from the original framework of the no-signaling principle.
Thus, the simulation paradigm points out that the inequality $ \sum_{a=\pm_{y}} P(a,+_y|A_1,B)\neq \sum_{a=\pm_{y}} P(a,+_y|A_2,B)$ can be realized. However, it is not related to the no-signaling principle; rather can be accounted for by the probabilistic nature of the simulation paradigm. This also resolves the consistency issue of $\cal PT$-symmetry theory.

\begin{remark}
The no-signaling principle is valid for the whole system governed by $\hat{H}_0\otimes I$ since this is a closed Hermitian system. See Ref. \cite{tang2016experimental} for more details on experimental investigation of the no-signalling principle in parity-time symmetric theory.
\end{remark}

\subsection{Further discussion}
Brody \cite{brody2016consistency} pointed out that for a finite dimensional closed system one cannot distinguish between $\cal PT$-symmetric and Hermitian Hamiltonians.
However, the inequality, $\sum_{a=\pm_{y}} P(a,+_y|A_1,B)\neq \sum_{a=\pm_{y}} P(a,+_y|A_2,B)$, shows how a distinction can be made in the sense of open systems.
As mentioned earlier, if we take $\rho=\eta^{-\frac{1}{2}}$ and $\rho'=\eta^{\frac{1}{2}}$ then
$H_0'=\eta^{\frac{1}{2}}H_0\eta^{-\frac{1}{2}}$ is Hermitian and $e^{-itH_0'}=\eta^{\frac{1}{2}}U_0(t)\eta^{-\frac{1}{2}}$ is unitary.
Hence one can prepare the Hermitian Hamiltonian $H_0'$ on $\mathbb C^2$ directly, without making use of the $\hat{H}_0$ on $\mathbb C^4$.
Thus, merely obtaining the probability distributions in $\mathbb C^2$, one does not know whether the subsystem is given by the simulation of $H_0$, or is obtained directly by the action of $H_0'$. In particular, Eq. (\ref{sig}) is valid in this case.
Note that now Alice's subsystem is equivalent to a closed system in $\mathbb C^2$ due to the fact that $e^{-itH_0'}$ is unitary.
Thus, the above discussion concludes that one cannot distinguish between $\cal PT$-symmetric Hamiltonian $H_0$ and the Hermitian Hamiltonian $H_0'$, for the closed subsystem in $\mathbb C^2$, as Brody suggested.
However, when Eq. (\ref{sig}) is violated in a certain simulation, Alice's subsystem must be viewed as an open system since such a violation does happen in a closed Hermitian system. And one can immediately learn from the violation that the subsystem is obtained by the simulation of $H_0$, rather than a direct action of the Hermitian Hamiltonian $H_0'$.
Thus, by utilizing Eq. (\ref{sig}), the simulation paradigm manifests how a distinction can be made between open $\cal PT$-symmetric and (closed) Hermitian Hamiltonians.

\section{conclusion}
In summary, we gave a full characterization of the embedding property and a simulation paradigm for the unbroken $\cal PT$-symmetric Hamiltonians in finite dimensional spaces.
This revealed how a general $\cal PT$-symmetric Hamiltonian can be realized, and provided a useful way to quantitatively analyse the traits of $\cal PT$-symmetric quantum systems. As an application, we found that the simulation paradigm can be employed to resolve the consistency problem and distinguish between open $\cal PT$-symmetric and Hermitian Hamiltonians.

\begin{acknowledgments}
This project is supported by National Natural Science Foundation of China (Grant No. 11571307).
\end{acknowledgments}


\begin{thebibliography}{99}
\bibitem{nonunitary1} M. E. Fisher, \href{https://journals.aps.org/prl/abstract/10.1103/PhysRevLett.40.1610}{Phys. Rev. Lett. {\bf 40}, 1610 (1978)}.

\bibitem{nonunitary2} J. L. Cardy, \href{https://journals.aps.org/prl/abstract/10.1103/PhysRevLett.54.1354}{Phys. Rev. Lett. {\bf 54}, 1345 (1985)}.

\bibitem{nonunitary3} J. L. Cardy and G. Mussardo, \href{https://www.sciencedirect.com/science/article/pii/0370269389908186}{Phys. Lett. B {\bf 225}, 275 (1989)}.

\bibitem{nonunitary4} A. B. Zamolodchikov, Nucl. Phys. B {\bf 348}, 619 (1991).

\bibitem{bender1998real} C. M. Bender and S. Boettcher, \href{https://journals.aps.org/prl/abstract/10.1103/PhysRevLett.80.5243}{Phys. Rev. Lett. {\bf 80}, 5243 (1998)}.

\bibitem{mostafazadeh2010pseudo} A. Mostafazadeh, Int. J. Geom. Methods Mod. Phys. {\bf 7}, 1191 (2010).

\bibitem{ruter2010observation} C. E. R\"{u}ter, K. G. Makris, R. El-Ganainy, D. N. Christodoulides, M. Segev, and D. Kip, Nat. Phys. {\bf 6}, 192 (2010).

\bibitem{chang2014parity} L. Chang, X. Jiang, S. Hua, C. Yang, J. Wen, L. Jiang, G. Li, G. Wang, and M. Xiao, Nat. Photonics {\bf 8}, 524 (2014).

\bibitem {PhysRevA.82.013629} E. M. Graefe, H. J. Korsch, and A. E. Niederle, \href {\doibase 10.1103/PhysRevA.82.013629} {Phys. Rev. A {\bf 82}, 013629 (2010)}.

\bibitem {PhysRevA.87.051601} M. Kreibich, J. Main, H. Cartarius, and G. Wunner, \href{\doibase 10.1103/PhysRevA.87.051601} {Phys. Rev. A {\bf 87}, 051601 (2013)}.

\bibitem{deffner2015jarzynski} S. Deffner and A. Saxena, \href{https://journals.aps.org/prl/abstract/10.1103/PhysRevLett.114.150601}{Phys. Rev. Lett. {\bf 114}, 150601 (2015)}.

\bibitem{bender17} C. M. Bender, D. C. Brody, and M. P. Muller, \href{https://journals.aps.org/prl/abstract/10.1103/PhysRevLett.118.130201}{Phys. Rev. Lett. {\bf 118}, 130201 (2017)}.

\bibitem{bender2007faster} C. M. Bender, D. C. Brody, H. F. Jones, and B. K. Meister, \href{https://journals.aps.org/prl/abstract/10.1103/PhysRevLett.98.040403}{Phys. Rev. Lett. {\bf 98}, 040403 (2007)}.

\bibitem{PhysRevA.78.042115} U. G\"{u}nther and B. F. Samsonov, Phys. Rev. A {\bf 78}, 042115 (2008).

\bibitem{bender2013pt} C. M. Bender, D. C. Brody, J. Caldeira, U. G\"{u}nther, B. K. Meister, and B. F. Samsonov, Philosophical Transactions of the Royal Society of London A: Mathematical, Physical and Engineering Sciences {\bf 371}, 20120160 (2013).

\bibitem{PhysRevA.90.054301} S.-L. Chen, G.-Y. Chen, and Y.-N. Chen, Phys. Rev. A {\bf 90}, 054301 (2014).

\bibitem{gunther2008naimark} U. G\"{u}nther and B. F. Samsonov, \href{https://journals.aps.org/prl/abstract/10.1103/PhysRevLett.101.230404}{Phys. Rev. Lett. {\bf 101}, 230404 (2008)}.

\bibitem{lee2014local} Y.-C. Lee, M.-H. Hsieh, S. T. Flammia, and R.-K. Lee, \href{https://journals.aps.org/prl/abstract/10.1103/PhysRevLett.112.130404}{Phys. Rev. Lett. {\bf 112}, 130404 (2014)}.

\bibitem{croke2015pt} S. Croke, \href{https://journals.aps.org/pra/abstract/10.1103/PhysRevA.91.052113}{Phys. Rev. A {\bf 91}, 052113 (2015)}.

\bibitem{brody2016consistency} D. C. Brody, \href{http://iopscience.iop.org/article/10.1088/1751-8113/49/10/10LT03/meta}{J. Phys. A: Math. Theor. {\bf 49}, 10LT03 (2016)}.

\bibitem{tang2016experimental} J.-S. Tang, Y.-T. Wang, S. Yu, {\it et al.}, Nat. Photonics {\bf 10}, 642 (2016).

\bibitem{uhlmann16} A. Uhlmann, Sci. China-Phys. Mech. Astron. {\bf 59}, 630301 (2016).

\bibitem{note0} In infinite dimensional case, $\cal P$ and $\cal T$ are defined by their actions on the dynamical variables, momentum operator $\hat{p}$ and the position operator $\hat{x}$, as follows. $\cal P$ reverses the sign of the dynamical variables: $\hat{p} \rightarrow -\hat{p}$ and $\hat{x} \rightarrow -\hat{x}$. Parity (space reflection) preserves the fundamental commutation relation $[\hat{x}, \hat{p}] = i$ (taking $\hbar =1$) in quantum mechanics. $\cal T$ has the effect $\hat{p} \rightarrow -\hat{p}$, $\hat{x} \rightarrow \hat{x}$, and $i \rightarrow -i$. There is change in the sign of $i$ because $\cal T$, like $\cal P$, is required to preserve the commutation relation $[\hat{x}, \hat{p}] = i$.

\bibitem{bender2007making} C. M. Bender, \href{http://iopscience.iop.org/article/10.1088/0034-4885/70/6/R03/meta}{Rep. Prog. Phys. {\bf 70}, 947 (2007)}.

\bibitem{deng2012general} J.-w. Deng, U. G\"{u}nther, and Q.-h. Wang, arXiv:1212.1861 (2012).

\bibitem{mannheim2013pt} P. D. Mannheim, Philosophical Transactions of the
Royal Society of London A: Mathematical, Physical
and Engineering Sciences {\bf 371}, 20120060 (2013).

\bibitem{horn2012matrix} R. A. Horn and C. R. Johnson, Matrix analysis (Cambridge university press, 2012).

\bibitem{bagarello2016non} F. Bagarello, J. Phys. A: Math. Theor. {\bf 49}, 215304 (2016).

\bibitem{gohberg1983matrices2} I. Gohberg, P. Lancaster, and L. Rodman, {\it Matrices
and indefinite scalar products} (Volume 8 of Operator
theory: Advances and applications, 1983).

\bibitem{note1} By Theorem \ref{thmheta}, there exists a $Q'$ such that
$H_3=Q'\Sigma Q'^{-1}$ and $Q'^\dag \eta Q'$ is a diagonal matrix.
Let us denote $Q'^{-1}Q$ by $C$. It then follows from $Q\Sigma Q^{-1}=Q'
\Sigma Q'^{-1}$ that $C\Sigma=\Sigma C$. Direct calculation shows
that $C$ is a diagonal matrix. Therefore, $Q^\dag \eta Q=C^\dag (Q'^\dag
\eta Q')C$ is also a diagonal matrix.

\bibitem{note2} In principle, quantum measurement theory allows us to utilize the von Neumann measurement $\{P_N, P_N^\perp\}$ to measure the state ${\cal U}_A\ket{\nu}$. The resulting state-ensemble of such a measurement is $\{P_N{\cal U}_A\ket{\nu}\bra{\nu}{\cal U}_A^\dag P_N, P_N^\perp{\cal U}_A\ket{\nu}\bra{\nu}{\cal U}_A^\dag P_N^\perp \}$. And ``post-selection''--a widely used theoretic and experimental technique in measurement--allows us to solely select the state $P_N{\cal U}_A\ket{\nu}\bra{\nu}{\cal U}_A^\dag P_N$ (unnormalised). The final state after post-selection, $P_N {\cal U}_A \ket{\nu}$ (pure and unnormalised), is a vector in the subspace $N$. Note that in the simulation paradigm, we may not begin with a pure state $\ket{\nu}$ but a general state $\sigma$ (trace one Hermitian operator). In this case, Eq. (\ref{67}) should be simply rewritten as $\sigma\mapsto \frac{ \rho'U(t)\rho \sigma \rho^\dag U(t)^\dag\rho'^\dag}{Tr (\rho'U(t)\rho \sigma \rho^\dag U(t)^\dag\rho'^\dag)}.$

\bibitem {reedsimon1980methods} M. Reed and B. Simon, Methods of modern mathematical physics: Vol 1, Functional analysis (Academic Press, 1980).

\end{thebibliography}
\end{document}